\documentclass[aps,preprint]{revtex4}%
\usepackage{amsfonts}
\usepackage{graphicx}
\usepackage{amsmath}
\usepackage{subfig}
\usepackage{amssymb}
\usepackage{caption}%
\providecommand{\U}[1]{\protect\rule{.1in}{.1in}}
\begin{document}
\title{Generation of $\chi^{2}$ solitons from the Airy wave through the parametric instability}
\author{Thawatchai Mayteevarunyoo}
\affiliation{Department of Telecommunication Engineering, Mahanakorn University of
Technology, Bangkok 10530, Thailand}
\author{Boris A. Malomed}
\affiliation{Department of Physical Electronics, School of Electrical Engineering, Faculty
of Engineering, Tel Aviv University, Tel Aviv 69978, Israel}

\begin{abstract}
Spontaneous creation of solitons in quadratic media by the downconversion,
i.e., parametric instability against the generation of fundamental-frequency
excitations, from the truncated Airy-wave (AW) mode in the second-harmonic
component is studied. Parameter regions are identified for the generation of
one, two, and three solitons, with additional small-amplitude
\textquotedblleft jets". Shares of the total power carried by individual
solitons are found. Also considered are soliton patterns generated by the
downconversion from a pair of AWs bending in opposite directions.

\end{abstract}
\maketitle

Transmission of waves with a specific intrinsic structure along curved paths
has drawn a great deal of interest in photonics and related areas. A
well-known class of self-bending beams is based on Airy wave (AW) packets, as
demonstrated, theoretically and experimentally, in quantum mechanics
\cite{Berry}, a variety of settings in optics \cite{Chr1}-\cite{Porras},
plasmonics \cite{plasm1}-\cite{plasm3}, electron beams \cite{el},
Bose-Einstein condensates \cite{BEC}, gas discharge \cite{discharge}, and
water waves \cite{water}. While Airy beams are eigenmodes of linear media,
their propagation in systems with cubic \cite{ChrSegev-nonlin}-\cite{Porras}
and quadratic (alias $\chi^{(2)}$) \cite{Ady-3wave}-\cite{Ady-3wave2},
\cite{Ady-Moti} nonlinearities was studied too. In the latter case, the
formation of two-color beams by the AW input launched in the
fundamental-frequency (FF) component (\textit{upconversion}) was analyzed
theoretically and demonstrated experimentally. The objective of the present
work is to analyze an alternative possibility of the \textit{downconversion},
i.e., the generation of solitons, that are fundamental eigenmodes of the
$\chi^{(2)}$ media, by the parametric instability (PI) of (truncated) Airy
beams launched in the second harmonic (SH), which are exact linear modes of
the system if FF perturbations are not added. In this connection, it is
relevant to note that formation (\textit{shedding}) of solitons by AWs in
cubic media was recently analyzed \cite{Marom} and observed experimentally
\cite{water}, while the formation of $\chi^{(2)}$ solitons via the PI of
straight SH wave packets, such as symmetric Gaussians, was studied still
earlier \cite{rev1}-\cite{rev3}.

It should be stressed that the scenario of the generation of solitons by small
random perturbations in the FF field interacting with the
parametrically-unstable SH input itself demonstrates instability, producing
quasi-random patterns including $\chi^{(2)}$ solitons and ``jets" (see below).
Nevertheless, systematic simulations make it possible to identify regular
features of the emerging soliton sets, such as the number of major solitons
and shares of the total power carried away by each of them.

The ``type-I" model of the one-dimensional $\chi^{(2)}$ system, with the
single FF component, $u$, and the SH component, $w$, is taken in the usual
scaled form \cite{rev1}-\cite{rev3}:%
\begin{gather}
iu_{z}=-(1/2)u_{xx}-u^{\ast}w,\nonumber\\
2iw_{z}=qw-(1/2)w_{xx}-(1/2)u^{2}, \label{type-I}%
\end{gather}
where $z$ and $x$ are the propagation distance and transverse coordinate, the
second derivatives represent the diffraction in the paraxial approximation,
$^{\ast}$ stands for the complex conjugate, and the mismatch parameter, $q$,
may be scaled to $q=+1,0,-1$. The system conserves the total power
(Manley-Rowe invariant) and momentum,%
\begin{equation}
P=\int_{-\infty}^{+\infty}\left(  |u|^{2}+4|w|^{2}\right)  dx,~M=i\int
_{-\infty}^{+\infty}\left(  u_{x}^{\ast}u+2w_{x}^{\ast}w\right)  dx, \label{W}%
\end{equation}
%
%H &=&\frac{1}{2}\int_{-\infty }^{+\infty }\left[
%|u_{x}|^{2}+|w_{x}|^{2}+2q|w|^{2}-\left( u^{2}w^{\ast }+\left( u^{\ast
%}\right) ^{2}w\right) \right] dx,  \label{H}
%\end{eqnarray}%
along with the corresponding Hamiltonian.

The self-bending linear modes for the SH in the absence of the FF component
are generated by the input in the form of truncated AWs \cite{Chr1},
\begin{equation}
w\left(  x,z=0\right)  =W_{0}\mathrm{Ai}\left(  \alpha x\right)  \exp\left(
\left(  \alpha/A\right)  x\right)  ,~u=0, \label{w0}%
\end{equation}
where $\mathrm{Ai}$ is the standard Airy function, $W_{0}$ and $1/\alpha$
determine the amplitude and spatial scale of the wave, while the truncation
parameter, $A$, makes the total power of input finite \cite{Chr1}:
\begin{equation}
P=P_{0}\alpha^{3}\sqrt{A/8\pi}\exp\left(  (2/3)A^{-2}\right)  ,~P_{0}\equiv
W_{0}^{2}/\alpha^{4}. \label{P}%
\end{equation}
%
%\begin{gather}
%w\left( x,z\right) =W_{0}\mathrm{Ai}\left( \alpha x-\frac{\alpha ^{4}}{16}%
%z^{2}+\frac{i}{2A}\alpha ^{3}z\right)   \notag \\
%\times \exp \left( -\frac{i}{2}qz-\frac{i}{96}\alpha ^{6}z^{3}+\frac{i}{4}%
%\alpha ^{3}xz\right)   \notag \\
%\times \exp \left( \frac{\alpha }{A}x-\frac{1}{8A}\alpha ^{5}z^{2}+\frac{i}{%
%4A^{2}}\alpha ^{4}z\right) .  \label{Ai}
%\end{gather}%

\begin{figure}[tb]
\centering\includegraphics[scale=0.5, bb=300 0 230 450]{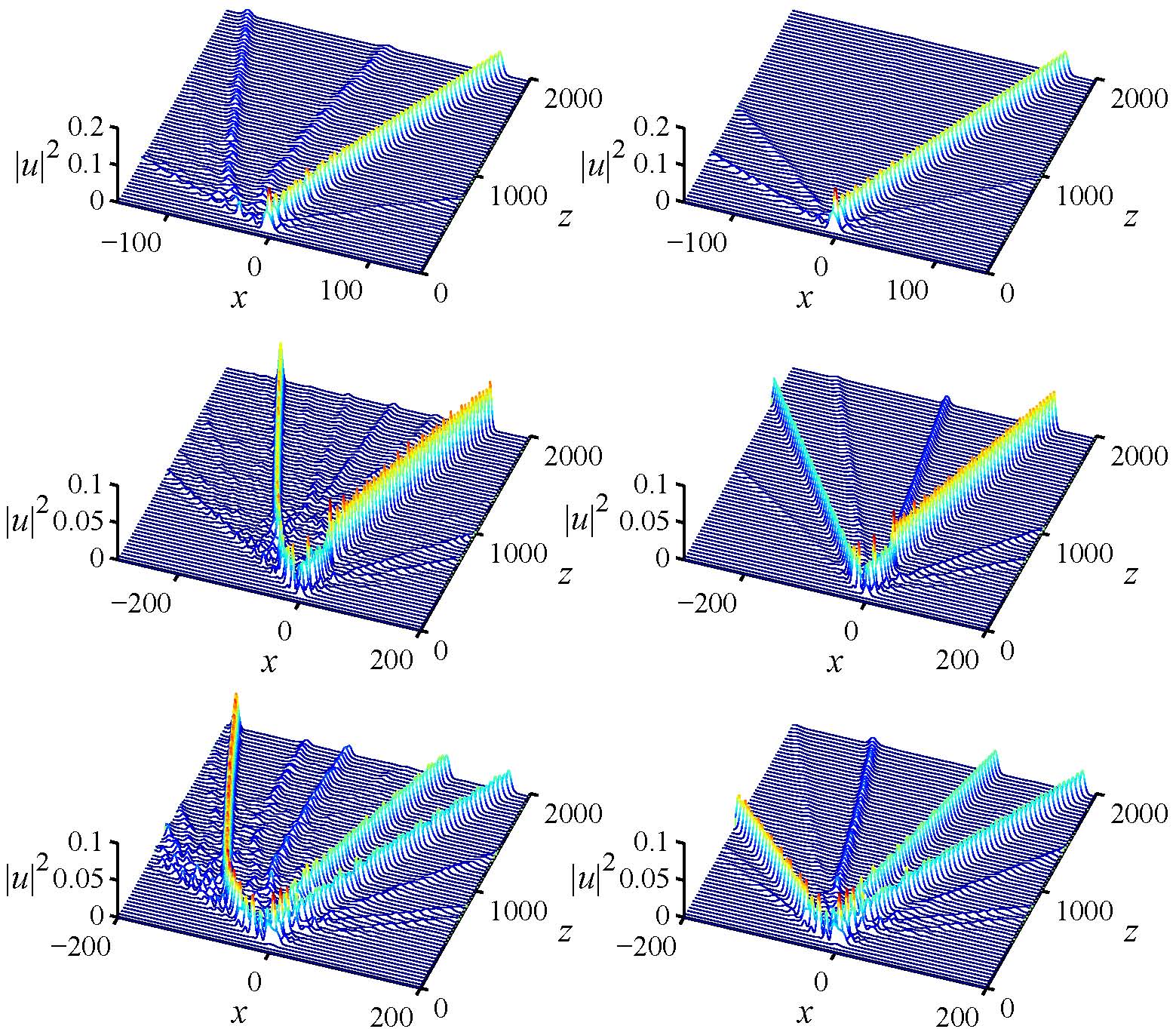}
\caption{(Color online) The top, middle, and bottom rows display the
generation of patterns featuring 1, 2, and 3 major $\chi^{(2)}$ solitons,
along with \textquotedblleft jets", from the SH input (\ref{w0}) with
$W_{0}=0.2571,\alpha=0.0855$; $W_{0}=0.2366,\alpha=0.0570$; and $W_{0}%
=0.2057,\alpha=0.0342$, respectively (in all cases, $A=10$). Here (and in Fig.
\ref{fig4} below) the intensity of the FF component is displayed; the
respective patterns of the SH field are shown in Figs. \ref{fig2} and
\ref{fig5}. To demonstrate how the coherence of the input affects the output,
left and right panels display, severally, the results generated by the full AW
input, and by its \textquotedblleft tailless" version, in which all but the
main lobe of the AW shape is cut off. This figure, as well as Figs. \ref{fig3}
and \ref{fig4}, \ref{fig5}, pertain to the system with zero mismatch, $q=0$.}%
\label{fig1}%
\end{figure}

\begin{figure}[tb]
\centering\includegraphics[scale=0.5, bb=300 0 230 450]{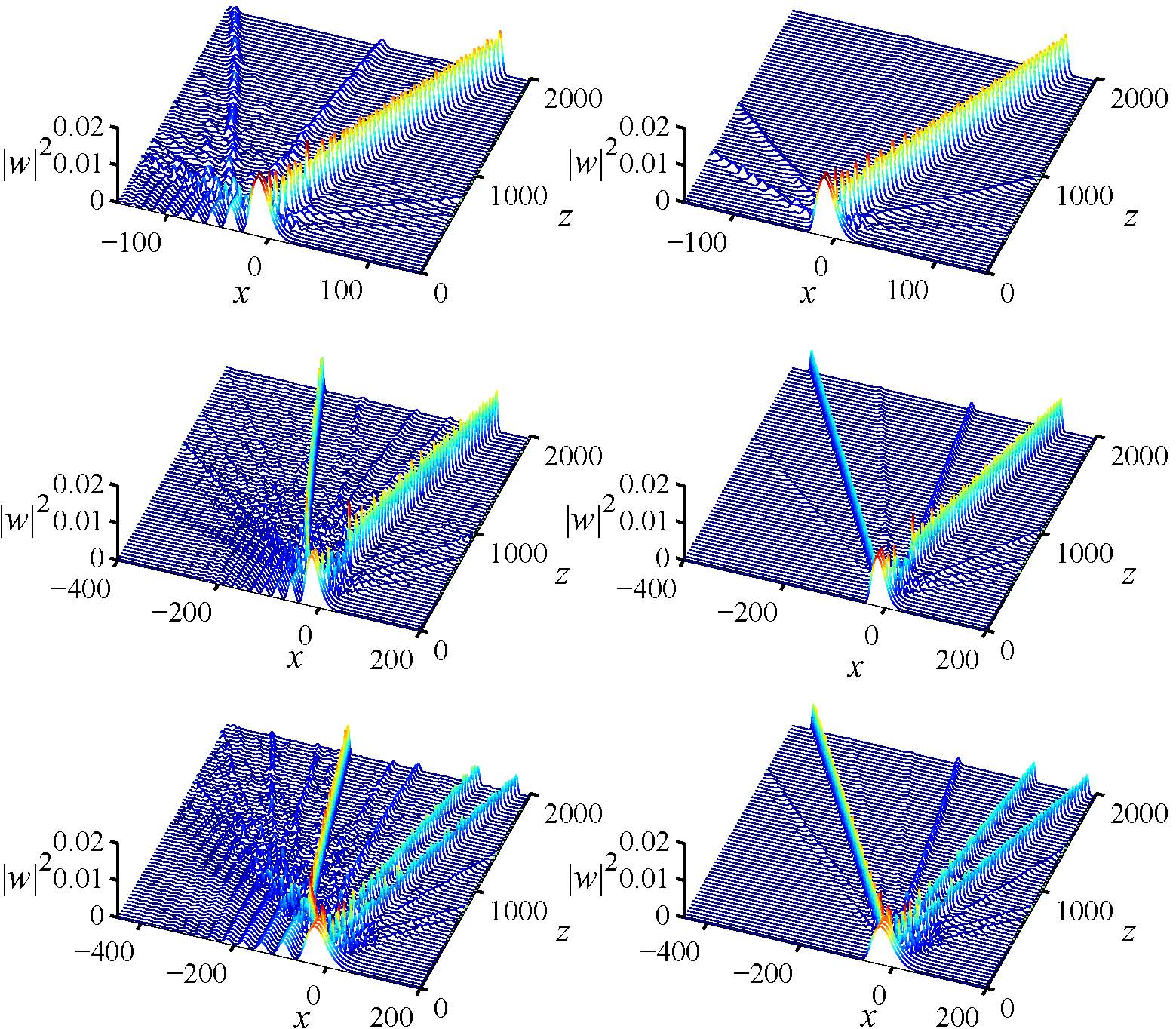}
\caption{(Color online) The intensity of the SH components of the patterns
displayed in Fig. \ref{fig1}.}%
\label{fig2}%
\end{figure}

Most results are displayed below for $A=10$, in which case the multi-lobe
structure of the truncated input seems virtually identical to the untruncated
one. Note that, in the case of $q=0$, which is chiefly considered here, the
scaling invariance of Eqs. (\ref{type-I}) demonstrates that, at fixed $A$, the
normalized total power of different inputs is actually measured by single
parameter $P_{0}$ defined in Eq. (\ref{P}).

In simulations, input (\ref{w0}) was used for the SH field, with small
(``seed") random perturbations in the FF mode initiating the onset of the PI
(the results do not essentially depend on characteristics of the seed). For
collecting the results which are displayed below, it was necessary to use a
large integration domain, of size $|x|\leq2^{11}$. In the course of the
simulations, it was checked that the total power is conserved, and the total
momentum remains equal to zero, see Eq. (\ref{W}).

As shown in the plots collected in the left column of Figs. \ref{fig1} and
\ref{fig2}, outcomes of the PI-initiated evolution can be summarized as the
generation of one or several clearly identified $\chi^{(2)}$ solitons,
supplemented by ``jets" with essentially smaller amplitudes, and still smaller
radiation features. Additional analysis demonstrates that the shape of
individual solitons is very close to that known in the standard $\chi^{(2)}$
model \cite{rev1}-\cite{rev3}. It is seen that the number of the major
solitons raises from $1$ to $2$ to $3$, as $P_{0}$ increases from $3164 $ to
$5294$ to $30925$. The generation of the patterns being initiated by random
small perturbations amplified by the PI, the picture always includes random
features, making it difficult to exactly identify values of $P_{0}$
corresponding to boundaries between the outcomes with different numbers of
major solitons. Nevertheless, the trend to the increase of the soliton number
with the growth of $P_{0}$ is evident. Results demonstrating more than three
major solitons are not presented here, as the respective simulations must be
run in a huge domain, and they would require a very wide waveguide for the
experimental realization.

If two or three solitons are generated by the AW input, their origin cannot be
unambiguously linked to particular lobes in the Airy-wave input, as the
emerging solitons become visible after some distance of ``latent" growth of
the PI, which separates the input and the multi-soliton pattern. Nevertheless,
the generation of the multi-soliton pattern, along with the ``jets", is a
coherent effect provided by the specific structure of the input. To further
address the latter point, plots collected in the right column of Fig.
\ref{fig1} report the results of the simulations performed, for the same
values of $W_{0}$, $\alpha$, and $A$, with the ``tailless" version of input
(\ref{w0}), in which only the main lobe of the Airy-wave profile is kept,
while the rest is eliminated. It is seen that the patterns generated by the
full AW, and solely by its main lobe, become essentially different with the
increase of the number of the generated solitons. Naturally, the patterns
originating from the ``tailless" inputs feature much weaker radiation
components. Nevertheless, the ``tailless" initial beam, that keeps the
inherent asymmetry of the AW, is basically different from the ordinary
symmetric inputs which were commonly used for the downconversion-driven
generation of $\chi^{(2)}$ solitons \cite{rev1}-\cite{rev3}.

\begin{figure}[tb]
\centering\includegraphics[scale=0.6,bb= 200 0 220 150]{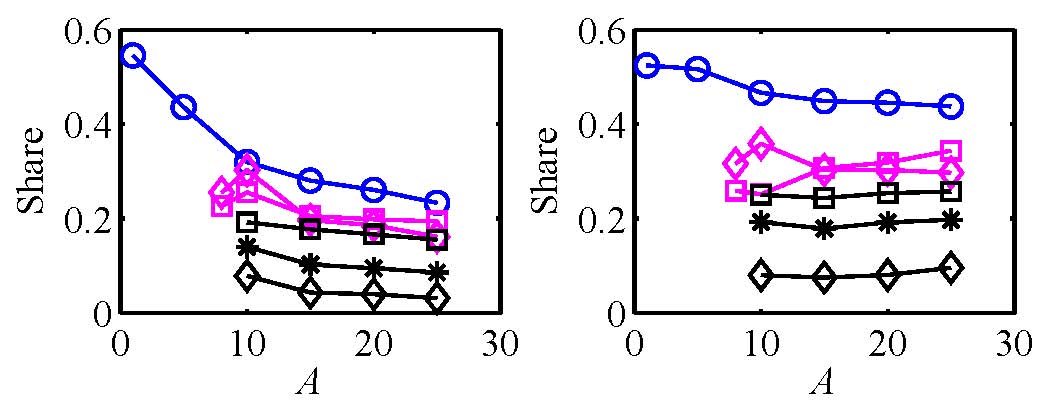}
\caption{(Color online) The share of the total power, defined per Eq.
(\ref{W}), which is carried away by each major $\chi^{(2)}$ soliton, vs. the
truncation factor, $A$, of the SH input (\ref{w0}). The number of lines of
each color corresponds to the number of major solitons, generated at the same
values of $W_{0}$ and $\alpha$ for which examples are displayed in Figs.
\ref{fig1} and \ref{fig2}. In the case when more than a single soliton is
generated, symbols $\mathbf{\square}$, ${\protect\LARGE \ast}$, and $\Diamond$
correspond to the leftmost, middle (in the case of three solitons), and
rightmost solitons, respectively. Like in Fig. \ref{fig1}, the left and right
panels pertain, severally, to the full and \textquotedblleft tailless"
inputs.}%
\label{fig3}%
\end{figure}

It is natural to measure the share of the total power carried by each major
soliton generated from the AW input (\ref{w0}). These shares are displayed, as
functions of the truncation factor $A$, in Fig. \ref{fig3}, for the full and
\textquotedblleft tailless" inputs alike. In particular, in the single-soliton
pattern generated by the full input, the share captured by the soliton
decreases roughly $\sim1/\sqrt{A}$ with the growth of $A$, which implies that
only a few leading lobes of the Airy-shaped input build the single soliton,
while the total power of the input grows $\sim\sqrt{A}$ [see Eq. (\ref{P})].
On the other hand, the combined share carried by all the solitons in the
multi-soliton pattern decreases essentially slower, which implies that, as
mentioned above, the generation of several solitons is a coherent effect
produced by the entire Airy-shaped input. Naturally, the result of the
evolution of the \textquotedblleft tailless" input is not sensitive to the
overall truncation. At large values of $A$, the net power share carried away
by the one/two or three solitons is larger, respectively, by factors
$\simeq1.8$ or $\simeq2.2$ in the case of the \textquotedblleft tailless"
initial conditions, as in that case much less power is spent on the generation
the jets and radiation. 

Note that, if a single soliton is generated, it always moves to the right,
i.e., in the same direction in which the input AW self-accelerates in the
framework of the linear equation for the SH. On the other hand, the two- and
three-soliton patterns feature, in Figs. \ref{fig1} and \ref{fig3}, a
situation in which a soliton with a larger amplitude may move in the opposite
direction, carrying a larger power than that carried to the right by one or
two other solitons. As mentioned above, in all cases the total momentum
(\ref{W}) remains equal to zero.

The spatial symmetry of the system suggests to consider a combination of two
inputs which launch the self-bending AWs in the opposite directions, with
separation $\xi>0$ and phase shift $\phi$ between them:%
\begin{gather}
w\left(  x,z=0\right)  =W_{0}\left[  \mathrm{Ai}\left(  \alpha\left(
x+\xi/2\right)  \right)  e^{i\phi/2+\alpha\left(  x+\xi/2\right)  /A}\right.
\nonumber\\
\left.  +\mathrm{Ai}\left(  \alpha\left(  -x+\xi/2\right)  \right)
e^{-i\phi/2\alpha\left(  -x+\xi/2\right)  /A}\right]  . \label{symm}%
\end{gather}
This input is subject to the Hermitian-symmetry condition, $w\left(
-x,z=0\right)  =w^{\ast}\left(  x,z=0\right)  $, which holds in the course of
the subsequent evolution. In the absence of the FF seed, input (\ref{symm})
gives rise to two linear SH waves passing through each other, while the
perturbed evolution creates $\chi^{(2)}$ solitons from the left and right AW
components, which interact with each other. Here we report results obtained
for large values of $\alpha$, corresponding to relatively small values of
$P_{0}$ in Eq. (\ref{P}), i.e., the generation of a single soliton by each AW
input separately (otherwise, the interaction pattern is very complex).
%Note also that large $\alpha $ gives
%rise to relatively broad solitons, helping to amplify interaction effects.

\begin{figure}[tb]
\centering\includegraphics[scale=0.5,bb=300 0 230 450]{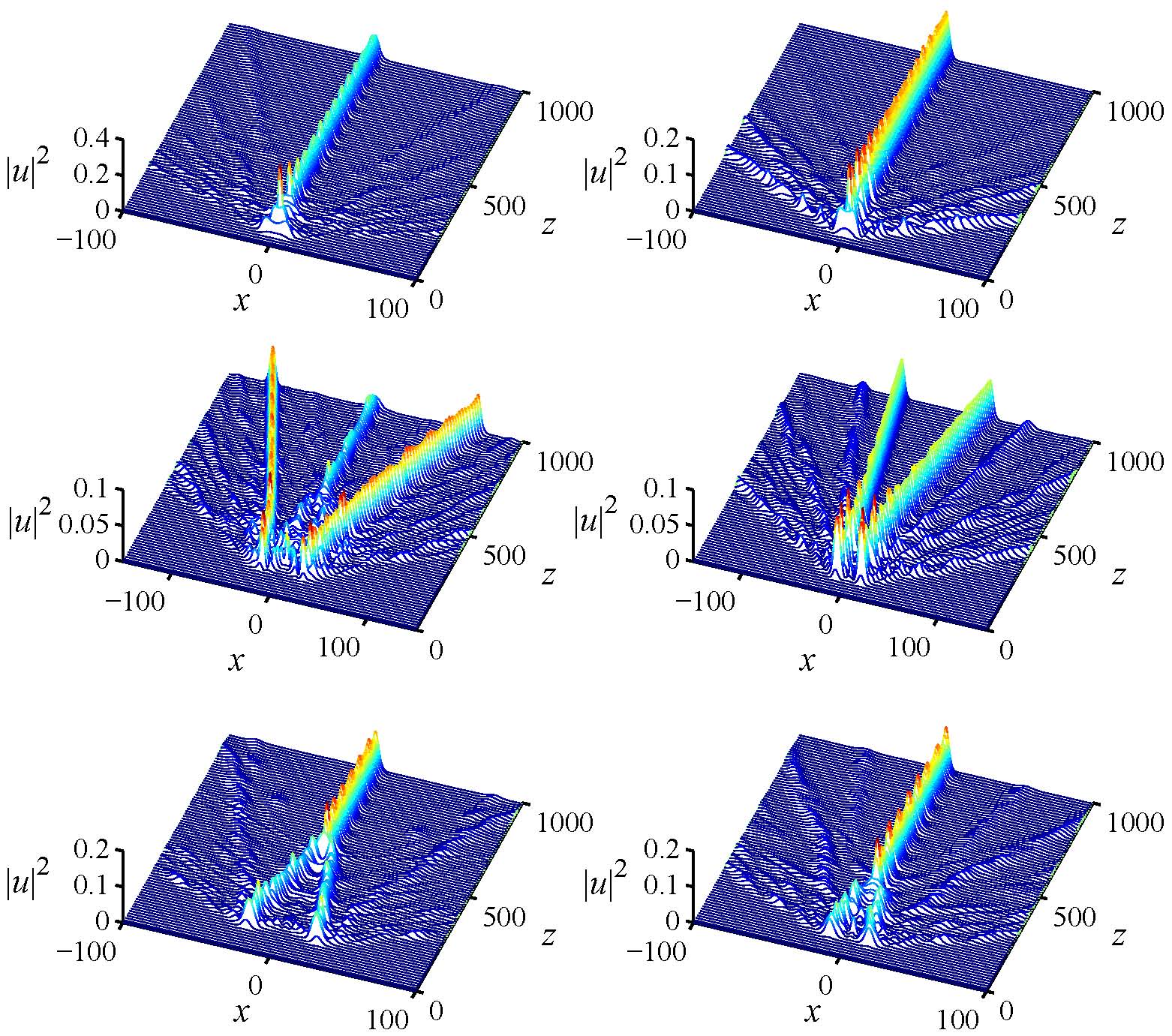}
\caption{(Color online) The left and right columns display, severally, the
generation of soliton patterns by the in- and out-of-phase pairs of the input
AWs in the SH, defined per Eq. (\ref{symm}) with $\phi=0$ and $\phi=\pi$. The
top, middle, and bottom rows demonstrate, respectively, the direct generation
of a single $\chi^{(2)}$ soliton (at $\xi=10$ for both $\phi=0$ and $\phi=\pi
$); the generation of two separating solitons with extra \textquotedblleft
jets" (at $\xi=37$ for $\phi=0$ and $\xi=20$ for $\phi=\pi$); and the fusion
of two emerging solitons into one (at $\xi=45$ for $\phi=0$ and $\xi=23$ for
$\phi=\pi$). In all cases, $W_{0}=0.309$, $\alpha=0.171$, $A=10$.}%
\label{fig4}%
\end{figure}

\begin{figure}[tb]
\centering\includegraphics[scale=0.5,bb=300 0 230 450]{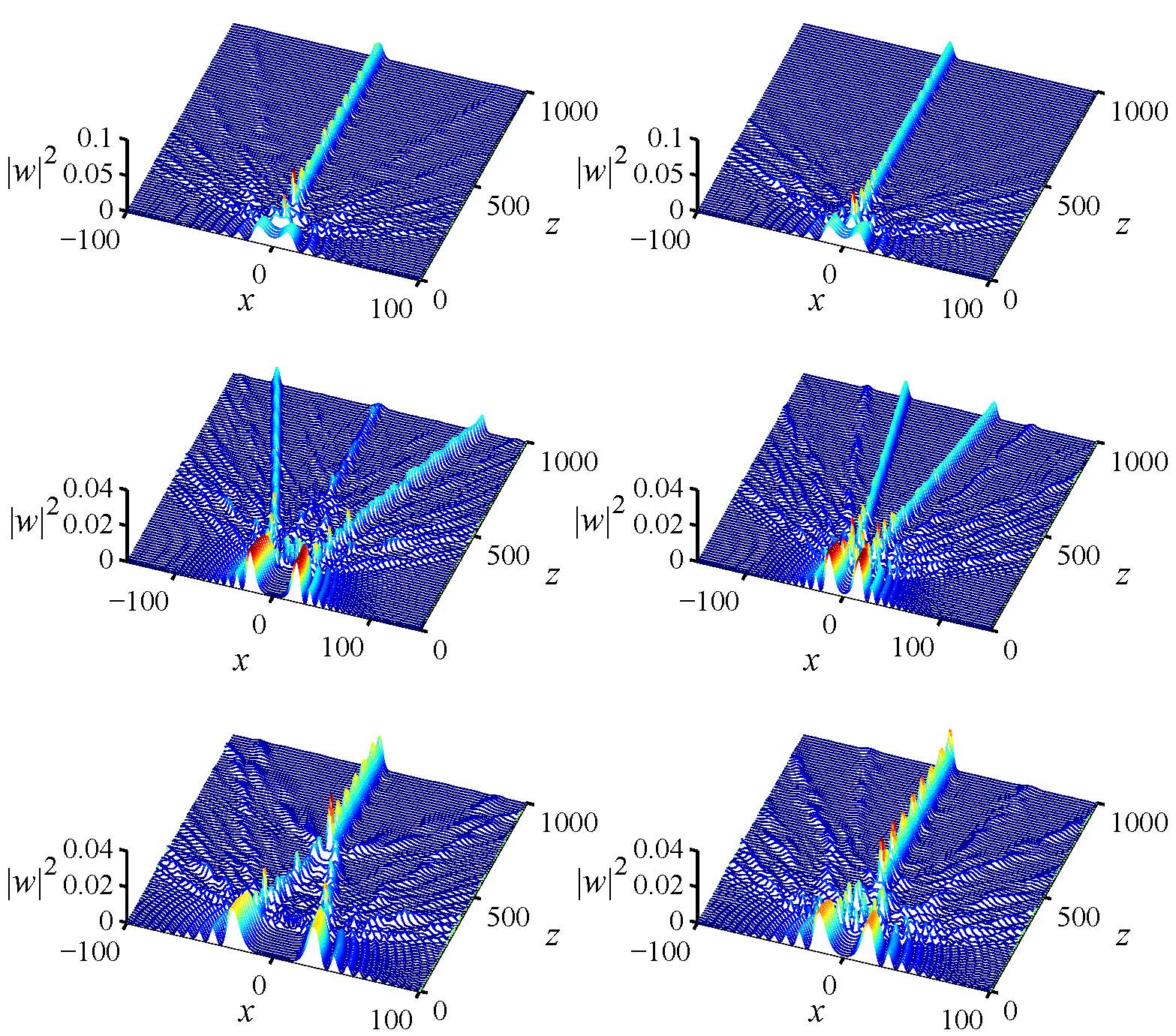}
\caption{(Color online) The intensity of the SH components of the patterns
displayed in Fig. \ref{fig4}.}%
\label{fig5}%
\end{figure}

Typical results of the interaction of two symmetric AW inputs are presented in
Figs. \ref{fig4} and \ref{fig5}, for both in-phase and out-of-phase pairs,
i.e., $\phi=0$ and $\phi=\pi$ in Eq. (\ref{symm}). When the separation between
them is small enough ($\xi\leq20$ for $\phi=0$, and $\xi\leq11$ for $\phi=\pi
$), the closely set left and right inputs generate a single soliton, in
agreement with the fact that each input carries a value of $P_{0} $ which is
far too small for the creation of more than one soliton. However, the increase
of $\xi$ allows the set of the two AW inputs to generate two solitons. In the
case of $\phi=\pi$, it is well known that the two solitons, placed at a
moderately large distance, repel each other \cite{KA}, hence they separate and
eventually appear as a pair of solitons moving in directions opposite to those
implied by the underlying AWs, as observed in Figs. \ref{fig4} and \ref{fig5}.

In the case of the in-phase input pair (\ref{symm}), with $\phi=0$, the direct
interaction between the emerging solitons should be attractive, on the
contrary to what is observed in the left middle panel of Fig. \ref{fig4},
which features separating solitons. However, it is seen too that the space
between the solitons is filled by radiation, the pressure from which may give
rise to repulsion \cite{cage1,cage2}. Note that, while the two solitons
separate, the radiation forms a relatively strong central ``jet". Additional
simulations demonstrate that, at specific values of $\xi$ (e.g., $\xi=38$ and
$\xi=17$, for $\phi=0$ and $\phi=\pi$, respectively) the central ``jet" may
form an additional soliton. Lastly, if $\xi$ is large enough ($\xi\geq40$ for
$\phi=0$ and $\xi\geq28$ for $\phi=\pi$), the two AW inputs generate two free
solitons, which then collide and merge into a single one, as clearly seen in
the bottom row of Fig. \ref{fig4}.

\begin{figure}[tb]
\centering\includegraphics[scale=0.6,bb= 200 0 220 150]{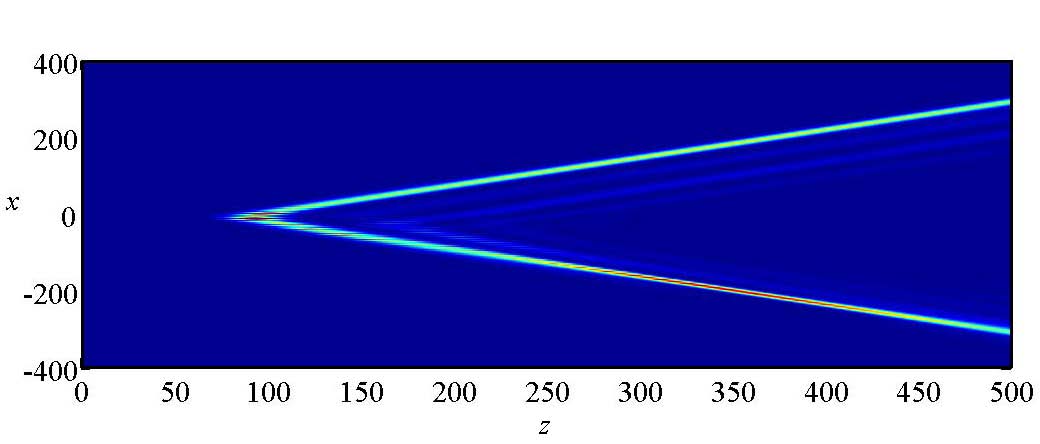}\caption{(Color
online) The top view of the generation of two solitons from SH input
(\ref{w0}) with $W_{0}=0.2571,\alpha=0.0855$, in the case of mismatch $q=+1$.}%
\label{fig6}%
\end{figure}

The effect of nonzero mismatch, $q=\pm1$ in Eq. (\ref{type-I}), was explored
too. While $q=-1$ prevents the generation of solitons, at the same values of
other parameters as considered above, $q=+1$ tends to increase the number of
the generated solitons. These trends are easily explained by the cascading
approximation, which eliminates the SH in favor of the FF component
\cite{rev1}-\cite{rev3}, $w\approx\left(  2q\right)  ^{-1}u^{2}$, and reduces
the $\chi^{(2)}$ system (\ref{type-I}) to the single nonlinear Schr\"{o}dinger
equation, $iu_{z}+(1/2)u_{xx}+\left(  2q\right)  ^{-1}|u|^{2}u=0$, which,
obviously, gives rise to one or multiple solitons at $q=+1$, and does not
produce them at $q=-1$. A typical example of the generation of two solitons at
$q=+1$ is displayed in Fig. \ref{fig6}, for parameter values corresponding to
$P_{0}=1237$ in Eq. (\ref{P}), which, as said above, is capable to generate
only a single soliton in the system with $q=0$.

%\section{Conclusion}

In conclusion, we have proposed a scenario for the interplay of AWs (Airy
waves) with the $\chi^{(2)}$ nonlinearity via downconversion, i.e., the PI
(parametric instability) seeded by small random FF perturbations added to the
truncated AW input in the SH component.
%This scenario is an alternative to
%the previously studied upconversion, with the AW launched in the FF
%component.
Although the downconversion mechanism, driven by the PI, exhibits inherent
randomness, it is possible to identify its basic features, such as the number
of emerging major solitons and power shares carried by them, including the
patterns generated by symmetric pairs of right- and left-bending AW inputs. A
natural generalization of the present analysis may be its development for
two-dimensional $\chi^{(2)}$ media, where AW inputs are available too, and
emerging solitons are stable \cite{rev2,rev3}.

\section*{Funding Information}

Thailand Research Fund (TRF) (RSA5780061).


\begin{thebibliography}{99}                                                                                               %


\bibitem {Berry}M. V. Berry and N. L. Balazs, Am. J. Phys. \textbf{47}, 264 (1979).

\bibitem {Chr1}G. A. Siviloglou and D. N. Christodoulides, Opt. Lett.
\textbf{99}, 213901 (2007).

\bibitem {Chr2}G. A. Siviloglou, J. Broky, A. Dogariu, and D. N.
Christodoulides, Phys. Rev. Lett. \textbf{99}, 213901 (2007).

\bibitem {Chr3}P. Polynkin, M. Kolesik, J. V. Moloney, G. A. Siviloglou, and
D. N. Christodoulides, Science 324, 229 (2009).

\bibitem {Ady-3wave}T. Ellenbogen, N. Voloch-Bloch, A. Ganany-Padowicz, and A.
Arie, Nature Phot. \textbf{3}, 395 (2009).

\bibitem {Tal}I. Dolev, T. Ellenbogen, and A. Arie, Opt. Lett. \textbf{35},
1581 (2010).

\bibitem {Ady-3wave2}I. Dolev and A. Arie, Appl. Phys. Lett. \textbf{97},
171102 (2010).

\bibitem {Chr4}R. El-Ganainy, K. G. Makris, M. A. Miri, D. N. Christodoulides,
and Z. Chen, Phys. Rev. A \textbf{84}, 023842 (2011).

\bibitem {ChrSegev-nonlin}I. Kaminer, M. Segev, and D. N. Christodoulides,
Phys. Rev. Lett. \textbf{106}, 213903 (2011).

\bibitem {Marom}Y. Fattal, A. Rudnick, and D. M. Marom, Opt. Exp. \textbf{19},
17298 (2011).

\bibitem {Marom2}A. Rudnick and D. M. Marom, Opt. Express. \textbf{19}, 25570 (2011).

\bibitem {nonlin-and-loss}A. Lotti, D. Faccio, A. Couairon, D. G. Papazoglou,
P. Panagiotopoulos, D. Abdollahpour, and S. Tzortzakis, Phys. Rev. A
\textbf{84}, 021807(R) (2011).

\bibitem {Morandotti}Y. Hu, Z. Sun, D. Bongiovanni, D. Song, C. Lou, J. Xu, Z.
Chen, and R. Morandotti, Opt. Lett. \textbf{37}, 3201 (2012).

\bibitem {Ady-Moti}I. Dolev, I. Kaminer, A. Shapira, M. Segev, and A. Arie,
Phys. Rev. Lett. \textbf{108}, 113803 (2012).

\bibitem {Denz}P. Rose, F. Diebel, M. Boguslawski, and C. Denz, Appl. Phys.
Lett. \textbf{102}, 101101 (2013).

\bibitem {Tsoy}I. M. Allayarov and E. N. Tsoy, ``Dynamics of Airy beams in
nonlinear media", Phys. Rev. A \textbf{90}, 023852 (2014).

\bibitem {inversion}R. Driben, Y. Hu, Z. Chen, B. A. Malomed, and R.
Morandotti, Opt. Lett. \textbf{38}, 2499 (2013).

\bibitem {Belic'}Y. Zhang, M. R. Beli\'{c}, H. Zheng, H. Chen, C. Li, Y. Li,
and Y. Zhang, Opt. Express. \textbf{22}, 7160 (2013).

\bibitem {Efremidis}N. K. Efremidis, Phys. Rev. A \textbf{98}, 023841 (2014).

\bibitem {Radik-VVK}R. Driben, V. V. Konotop, and T. Meier, Opt. Lett.
\textbf{39}, 5523 (2014).

\bibitem {Porras}C. Ruiz-Jim\'{e}nez, K. Z. N\'{o}brega, and M. A. Porras,
Opt. Express. \textbf{23}, 8918 (2015).

\bibitem {plasm1}A. Minovich, A. E. Klein, N. Janunts, T. Pertsch, D. N.
Neshev, and Y. S. Kivshar, Phys. Rev. Lett. \textbf{107}, 116802 (2011).

\bibitem {plasm2}L. Li, T. Li, S. M. Wang, C. Zhang, and S. N. Zhu, Phys. Rev.
Lett. \textbf{107}, 126804 (2011).

\bibitem {plasm3}I. Epstein and A. Arie, Phys. Rev. Lett. \textbf{112}, 023903 (2014).

\bibitem {el}N. Voloch-Bloch, Y. Lereah, Y. Lilach, A. Gover, and A. Arie,
Nature \textbf{494}, 331 (2013).

\bibitem {BEC}N. K. Efremidis, V. Paltoglou, and W. von Klitzig, Phys. Rev. A
\textbf{87}, 043637 (2013).

\bibitem {discharge}M. Clerici, Y. Hu, P. Lassonde, C. Mili\'{a}n, A.
Couairon, D. N. Christodoulides, Z. Chen, L. Razzari, F. Vidal, F.
L\'{e}gar\'{e}, D. Faccio, R. Morandotti, Sci. Adv. \textbf{1}, e140011 (2015).

\bibitem {water}S. Fu, Y. Tsur, J. Zhou, L. Shemer, and A. Arie, Phys. Rev.
Lett. \textbf{115}, 034501 (2015).

\bibitem {rev1}G. I. Stegeman, D. J. Hagan, and L. Torner, Opt. Quant.
Electron. \textbf{28}, 1691-1740 (1996).

\bibitem {rev2}C. Etrich, F. Lederer, B. A. Malomed, T. Peschel, and U.
Peschel, Prog. Opt. \textbf{41}, 483 (2000).

\bibitem {rev3}A. V. Buryak, P. Di Trapani, D. V. Skryabin, and S. Trillo,
Phys. Rep. \textbf{370}, 63 (2002).

\bibitem {KA}Y. S. Kivshar and G. P. Agrawal, \textit{Optical Solitons: From
Fibers to Photonic Crystals} (Academic Press, San Diego, 2003).

\bibitem {cage1}R. Driben, A. V. Yulin, A. Efimov, and B. A. Malomed, Opt.
Express. \textbf{21}, 19091 (2013).

\bibitem {cage2}S. F. Wang, A. Mussot, M. Conforti, X. L. Zeng, and A.
Kudlinski, Opt. Lett. \textbf{40}, 3320 (2015).
\end{thebibliography}
\end{document}